\documentstyle[prb,aps]{revtex}
\twocolumn
\begin{document}
\title{Comment on ``$c$-axis Josephson tunneling in $d_{x^2-y^2}$-wave superconductors''}
\author{G. B. Arnold$^*$, R. A. Klemm$^{\dag}$, W. K{\"o}rner$^{\ddag}$, K. Scharnberg$^{\ddag}$}
\address{$^*$Physics Department, University of Notre Dame, Notre Dame, Indiana 46556}

\address{$^{\dag}$Max-Planck-Institut f{\"u}r Physik komplexer Systeme, N{\"o}thnitzer Str. 38, D-01187
Dresden, Germany }

\address{$^{\ddag}$I. Institut f{\"u}r Theoretische Physik, Universit{\"a}t Hamburg, Jungiusstr.
9, D-05535 Hamburg, Germany }
\maketitle

Recently, Maki and Haas (MH) presented
an expression, Eq. (3), for the critical current $I_c$ for coherent $c$-axis 
tunneling across the interface between identical
$d_{x^2-y^2}$-wave superconductors twisted a relative angle
{$\alpha$} about the $c$-axis.\cite{MakiHaas}  They implied that 
Eq. (3) was valid to all orders in the $c$-axis tunneling
probability $t_{\perp}$, accounting for ``the effects of Andreev reflections
at the grain boundary".\cite{MakiHaas}  However, they made many qualitative and quantitative errors, which invalidate their
conclusions.

The errors are present in their Eq. (3).  MH
state that this result is obtained ``by applying the
Ambegaokar-Baratoff formalism", but a derivation is neither given nor
 cited.  The result of Ambegaokar and
Baratoff (AB) is based on the tunneling Hamiltonian
approximation,\cite{AB} valid to lowest order in 
$t_{\perp}$, while the MH result clearly involves higher orders in
$t_{\perp}$. Therefore, we first take this equation at face value
and examine its consequences.

 MH assumed a 
cylindrical, two-dimensional Fermi surface in each 
superconductor adjacent to the interface,
 in contrast to the experimental tight-binding Fermi surface of
Bi$_2$Sr$_2$CaCu$_2$O$_{8+\delta}$  (Bi2212). 
In the s-wave limit,  the MH Eq. (3) for  $I_c$ is independent
 of $t_{\perp}$, which  is
clearly unphysical. Even if  Eq. (3) were to apply for
$I_cR/e$, it is wrong, because exact calculations for s-wave
superconductors
 have shown that $I_cR/e$ for an $SIS$ junction depends upon
$t_{\perp}$.\cite{Arnold}  At low temperature $T$, 
 $I_cR/e$ increases from the AB value to twice that
value  as $t_{\perp}$ increases from 0 to unity. Only as 
 $T\rightarrow T_c$ does $I_cR/e$ become independent
of $t_{\perp}$.

In addition, the MH result disagrees with an exact calculation of
$I_c$ for tunneling between $SIS$ layered superconductors obtained
 by Arnold and Klemm (AK),\cite{ArnoldKlemm} given in
their Eq. (18). Comparing the MH Eq. (3) with the AK Eq. (18), we
identify many significant differences.  The exact   AK Eq. (18) contains 
the phase difference $\delta\phi$ across the tunnel barrier in the coefficient
of each order in 
$t_{\perp}$, 
 and $I_c$ is
optimized for each $\alpha$ by setting $\delta\phi=\delta\phi^*<\pi/2$. 
 Each Andreev reflection should also
 involve $\delta\phi$, but the MH Eq. (3) evidently  set
$\delta\phi^*=\pi/2$. 
The MH gap $\Delta_0$ is evidently independent of $T$. This
disagrees with the AB limit.  In the d-wave case, as
$t_{\perp}\rightarrow 0$, the MH result also disagrees with the d-wave AB
limit.

Andreev reflections at the interface could only arise if it were an $SNS$
junction. It would then be  much stronger than the intrinsic
interplanar couplings in the single crystals, which are generally agreed to be
of the $SIS$ type.  However, MH  neglected altogether the intrinsic  
tunneling between the junctions in the layered superconducting half-spaces
adjacent to the interface.\cite{ArnoldKlemm}   Furthermore, the
$t_{\perp}({\bf k})$  employed by MH in their 
Eq. (6) is inappropriate for the cylindrical Fermi surface 
they employed. \cite{MakiHaas}  As
shown previously, 
\cite{ArnoldKlemm,BKS} tight-binding Fermi surfaces of layered
superconductors such as Bi2212 misoriented  relative to each other
 also introduce a
strong $\alpha$ dependence of $I_c$ for coherent
tunneling, even in the s-wave case, especially with a $t_{\perp}({\bf k})$
analogous to that used in Eq. (6) of MH.\cite{ArnoldKlemm} 

The $I_c$s observed by Li {\it et al.} are equal to those observed in
single crystals,\cite{Li} which are not ``small", as claimed by
MH.  Li {\it et al.} fed the c-axis current into the top of
the crystal. We have studied this case theoretically, \cite{Koerner} and find
that within a few layers from the current leads, the current
becomes uniform over the crystal surface on the scale of the
c-axis penetration depth, $\lambda_c>100\mu$m.
 However, in the
whisker experiment of Takano {\it et al.},\cite{Takano} the current
 was fed in from the edge,
producing current inhomogeneities on the scale of the Josephson
penetration depth, $\approx 2-3\mu$m.\cite{Koerner}

Finally, we note that the $\alpha$ dependence of $I_c$ observed in
the experiments of Takano {\it et al.} may have an
extrinsic origin, due to  systematic variations with $\alpha$ of the
Bi2212-Bi$_2$Sr$_2$Ca$_2$Cu$_3$O$_{10+\delta}$ mixture and additional
insulating barrier thicknesses near the
 cross-whisker interface.\cite{Klemm,Singapore}

\end{document}